\documentclass[preprint,aps,prd,showpacs,nofootinbib]{revtex4}
\usepackage[colorlinks, linkcolor=blue, citecolor=red, urlcolor=red]{hyperref}
\usepackage{mathrsfs}
\usepackage{amsmath}
\usepackage{graphicx}
\usepackage{subfigure}
\def\be{\begin{equation}}
\def\ee{\end{equation}}
\def\ba{\begin{eqnarray}}
\def\ea{\end{eqnarray}}

\renewcommand{\[}{\left[}
\renewcommand{\]}{\right]}

\def\be{\begin{equation}}
\def\ee{\end{equation}}
\def\ba{\begin{eqnarray}}
\def\ea{\end{eqnarray}}

\begin{document}

\title{Spherical collapse in the extended quintessence cosmological models}

\author{Yize Fan$^{1}$, Puxun Wu$^{2, 3}$ and  Hongwei Yu$^{1, 2,}$\footnote{Corresponding author} }
\affiliation{$^1$Institute of Physics and Key Laboratory of Low
Dimensional Quantum Structures and Quantum
Control of Ministry of Education,\\
Hunan Normal University, Changsha, Hunan 410081, China \\
$^2$Center for Nonlinear Science and Department of Physics, Ningbo
University, Ningbo, Zhejiang 315211, China
\\$^3$Center for High Energy Physics, Peking University, Beijing 100080, China
}

\begin{abstract}
We  use the spherical collapse method to investigate the nonlinear density perturbations of pressureless matter in the cosmological models with the extended quintessence as dark energy in the metric and Palatini formalisms.   We find that for both formalisms, when the coupling constant is negative, the deviation  from the $\Lambda$CDM  model is the least according to the evolutionary curves of  the linear density contrast $\delta_{c}$  and virial overdensity $\Delta_{v}$, and it is less than $1\%$. And this indicates that,  in the extended quintessence cosmological models in which the coupling constant is negative, all quantities dependent  on $\delta_{c}$ or $\Delta_{v}$ are essentially  unaffected if the linear density contrast or the virial overdensity of the $\Lambda$CDM model is used as an approximation. Moreover, we find  that  the differences between different formalisms are very small in terms of structure formation, and  thus cannot be used to distinguish the metric and Palatini formalisms.
\end{abstract}

\maketitle

\section{Introduction}\label{sec1}
Overwhelming evidence has indicated  that our Universe is   undergoing an accelerated expansion, and this surprising
observation can   be accounted for by a modification of  the theory of general relativity on the cosmic scale (see Refs.~\cite{Sotiriou,Felice} for recent reviews),  or the existence of  an exotic energy component with negative pressure, called dark energy  (see Ref.~\cite{DE} for recent reviews).    Popular candidates for dark energy include the cosmological constant~\cite{Padamanabhan, Caldwell, Parker} and some  dynamical scalar fields, such as quintessence~\cite{Quintessence}, phantom~\cite{Phantom} and quintom~\cite{Quintom}, etc., which couple minimally with  gravity.   More generally,  one may also assume a nonminimal coupling between the scalar field dark energy and gravity, which  can be  generated naturally when quantum corrections are considered and are essential for the renormalizability of the scalar field theory in curved space.   When a normal scalar field is assumed to couple with the curvature scalar defined in the metric formalism, the resulting dark energy  is called extended quintessence~\cite{Sahni,  Uzan,       Elizalde, Bartolo,  Faraoni, Chiba, Holden,  Abbott}, and it is in fact a special case of the scalar-tensor theories~\cite{Hwang, Hwang2, Schimd, Riazuelo,    Boisseau, Pettorino1, Pettorino2,   Sadjadi,   Hrycyna, Amendola, Baccigalupi, Bean, Esposito}. Recently,  by replacing the curvature scalar defined in the metric formalism with that in the Palatini formalism, we proposed a new extended quintessence in Ref.~\cite{Wang2013}.
It is worth noting that, in the general relativity limit,  two different metric formalisms give the same field equations~\cite{Misner}. However, once gravity is modified, they lead to different results.

In our previous work~\cite{Fan2015}, we studied the linear cosmological perturbations of the extended quintessence in both the metric and Palatini formalisms. We found that  in the Palatini formalism  the value of the gravitational potential of the extended quintessence  is always smaller than that of the minimal coupling case.  This is different from what happens in the metric formalism where the gravitational potential is less than that in the minimal coupling case when  the coupling constant is positive, while it is larger if the coupling constant is negative.   The expressions of the effective Newton's constant in the metric and Palatini formalisms are different. A negative coupling constant enhances the gravitational interaction, while a positive one weakens it.  We also found that  the metric and Palatini formalisms give different linear growth rates, but the difference is very small, and current observation cannot distinguish them effectively.

In the present work, we plan to investigate the nonlinear structure formation of the extended quintessence in the metric and Palatini formalisms by using the spherical collapse method. Spherical collapse is a semianalytical  method often used to  evolve structures by following perturbations into the nonlinear region~\cite{Peebles67, Gunn72, Padmanabhan93, Bilic04, Pace10, Popolo}.  In this method, perturbations are assumed to be  spherically symmetric nonrotating objects. Since perturbations will be overdense under the attractive action of  gravity,  the spherically symmetric objects will decouple from the background Hubble expansion, reach a point of maximum expansion (turnaround) and finally collapse  to a singularity.  However, the final singular phase does not happen, since  the kinetic energy associated with the collapse is converted into random motions, creating a virialized equilibrium  structure.  The spherical collapse method has been used in different cosmological models, including $\Lambda$CDM~\cite{Engineer00, Padmanabhan93}, quintessence~\cite{Wang98, Mainini},  quintessence with an interaction between it and dark matter~\cite{Baccigalupi2002}, extended quintessence~\cite{Pace10, Pace2009}, K-essence~\cite{Mann}, and so on.  It has been found that  the spherical collapse model is very successful in reproducing results of N-body simulations~\cite{Pace10,Pace2009, Hiotelis}, which are the best ways to investigate the structure formation.

Besides the extended quintessence, we  also consider the $\Lambda$CDM and minimally coupled quintessence models. Detailed comparisons among these models are given, including the equation of state of dark energy,  the evolutions of the energy density of dark energy and the Hubble parameter, and the nonlinear structure formation. The paper is organized as follows: In Sec. II, we show the  background cosmological models considered here. The  perturbed equations are studied in Sec. III, and the spherical collapses  are analyzed in Sec. IV. We conclude in Sec. V. Throughout this paper, unless specified, we adopt the metric signature ($-, +, +, +$). Latin indices run from $0$ to $3$, and the Einstein convention is assumed for a repeated index.

\section{Background Cosmology}

\subsection{Einstein$-$de Sitter and $\Lambda$CDM models}
The simplest candidate for dark energy is the cosmological constant $\Lambda$, whose equation of state  is always a constant as $w=-1$. The cosmological model consisting of $\Lambda$ and cold dark matter (CDM), named $\Lambda$CDM,  describes the present cosmic  expansion well. Its Friedmann equation has the form
\be\label{H}
H^{2}{(z)}=H_{0}^{2}\left[\Omega_{m0}(1+z)^{3}+(1-\Omega_{m0})\right] \ee
  in a spatially flat, homogenous and isotropic universe  described by the Friedmann-Lemaitre-Robertson-Walker (FLRW) metric $g_{\mu\nu}=diag\{-1, a^{2}(t), a^{2}(t), a^{2}(t)\}$, with $a$ being the scale factor and $t$ the cosmic time. Here, $z$ is the redshift,  $H=\frac{\dot{a}}{a}$ is the Hubble parameter,  and $H_{0}$ means its present value. Throughout this paper, a dot represents a derivative with respect to the cosmic time.  $\Omega_{m}$ represents the  matter density normalized to the
critical density $\rho_{crit}=3H^{2}/8\pi G$, and $\Omega_{m0}$ is its present value. When  $\Omega_{m}=1$, the $\Lambda$CDM model reduces to the Einstein$-$de Sitter (EdS) one.

\subsection{Minimally and nonminimally coupled quintessence models}

The action for the nonminimally coupled quintessence model, which has been under extensive study in recent years~\cite{Baccigalupi2002}, takes the  following form  when the radiation is neglected:
\be\label{action1}
S=\int d^4x\sqrt{-g}\[\frac{1}{2\kappa^2}f(\varphi,\hat{R})+{\mathcal{L}_{\varphi}}+{\mathcal{L}_{m}}\],\;\ee
where $g$ is the determinant of the background metric $g_{\mu\nu}$, $\kappa^2=8\pi G$, $G$ is the Newton gravitational constant, $\hat R$ is the Ricci scalar defined as
$\hat{R}=g^{\mu\nu} \hat{R}_{\mu\nu}=g^{\mu\nu}( \Gamma^\alpha_{\mu\nu,\alpha}-\Gamma^\alpha_{\mu\alpha,\nu}
+\Gamma^\alpha_{\alpha\lambda}\Gamma^\lambda_{\mu\nu}-\Gamma^\alpha_{\mu\lambda}\Gamma^\lambda_{\alpha\nu})$,  and $\Gamma^\lambda_{\mu\nu}$ is the connection.  In the metric formalism, only the metric is a variable and the connection is the Levi-Civit\`{a} type, while in the Palatini formalism, both the metric and the connection are independent variables.
 ${\mathcal{L}_{\varphi}}$ and ${\mathcal{L}_{m}}$ stand for the Lagrangians of  the scalar field and pressureless matter, respectively. ${\mathcal{L}_{\varphi}}$ has the form
\be\label{c3}
\mathcal{L}_{\varphi} =-\frac{1}{2}\nabla^{\mu}\varphi\nabla_{\mu}\varphi-V(\varphi).\;\ee
Here, $\nabla_{\mu}$ is the usual covariant derivative with respect to the Levi-Civit\`{a} connection, and $V(\varphi)$ is the potential of the scalar field, which is assumed to be  of an exponential form
\be\label{c4}
V(\varphi)=V_{0} e^{\lambda  \kappa \varphi},\;\ee
where $V_{0}$ is a nonzero constant and   the constant $\lambda$  is set as $\lambda=-1.1$ in our analysis.

In the action given in Eq.~(\ref{action1}), $f$ is an arbitrary function of $\varphi$ and $\hat{R}$. In the present paper, we take $f(\varphi,\hat{R})$ to  be of the following form:
\be\label{c5}
f(\varphi,\hat{R})=F(\varphi)\hat R, \quad F(\varphi)=1+\chi \kappa^2\varphi^2,\;\ee
where $\chi$ is the coupling constant and $\chi=0$ corresponds to the minimally coupled case. In this case,  we find that the Ricci scalar has the form $ \hat R=R+ \xi (\frac{3}{2}\frac{\nabla_\sigma F\nabla^\sigma F}{F^2}
-3\frac{\nabla_\sigma\nabla^\sigma F}{F}) $
  with  $R$ being that defined in the metric formalism (see the Appendix  of Ref.~\cite{Fan2015} for details). Here, $\xi=0$ or $1$.  $\xi=0$ corresponds to the metric formalism and $\xi=1$ to the Palatini formalism.

Varying the action with respect to the metric $g_{\mu\nu}$ gives the field equations~\cite{Wang2013,Fan2015}
   \be\label{c6} F \hat{ R}_{\mu\nu}-\frac{1}{2}F\hat R g_{\mu\nu}-(1-\xi)({\nabla_\mu\nabla_\nu F}
-g_{\mu\nu}\nabla_{\sigma}\nabla^{\sigma} F) = \kappa^2\bigg[T^{(\varphi)}_{\mu\nu}+T^{(m)}_{\mu\nu}\bigg].\;\ee
One can see that when $\chi=0$,   $F=1$ and Eq.~(\ref{c6}) gives the field equation of the minimally coupled quintessence, which indicates that the metric and Palatini formalisms give the same result in the limit of general relativity.

In Eq.~(\ref{c6}),  $T^{(m)}_{\mu\nu}$ and  $T^{(\varphi)}_{\mu\nu}$ are the energy-momentum tensors of  pressureless  matter and  the scalar field, respectively, and  are defined as
\be\label{c7} T^{(m)}_{\mu\nu}=-\frac{2}{\sqrt{-g}}\frac{\delta(\sqrt{-g}\mathcal{L}_{m})}{\delta g^{\mu\nu}}\;\ee
and
\be\label{c8} T^{(\varphi)}_{\mu\nu}=\nabla_{\mu} \varphi \nabla_{\nu} \varphi-g_{\mu\nu} \[\frac{1}{2}(\nabla^{\alpha}\varphi \nabla_{\alpha}\varphi)+V\].\;\ee

Assuming that our Universe is spatially flat, homogeneous and isotropic, and  is described by the FLRW metric, one can obtain the Friedmann equations from Eq.~(\ref{c6})~\cite{Faraoni, Chiba, Fan2015, Esposito, Wang2013}:
\be\label{c9} H^2=\frac{\kappa^2}{3}\bar{\rho}_m+\frac{\kappa^2}{3}\bigg(\frac{1}{2}{\dot\varphi}^2+V\bigg)
-H\dot F- \chi \kappa^2H^2\varphi^2 -\xi \frac{\dot F^2}{4F},\;\ee
\be\label{c10} -3H^2-2\dot{H}=\kappa^2\bigg(\frac{1}{2}\dot\varphi^2-V\bigg)
+2H\dot{F}+\ddot{F}+ \chi  \kappa^2 \varphi^2(3H^2+2\dot{H})-\xi \frac{3\dot F^2}{4F}.\;\ee
Here,  $\bar{\rho}_{m}$ means the background energy density of pressureless matter.   Rewriting the above two equations in the standard form, one can define the effective energy density and the effective pressure of dark energy:
\be\label{c11} \rho_{\varphi,eff}=\frac{1}{\kappa^2}\left[\kappa^2\bigg(\frac{1}{2}{\dot\varphi}^2+V\bigg)
-3H\dot F-3 \chi \kappa^2H^2\varphi^2 -\xi \frac{3\dot F^2}{4F}\right],\;\ee
\be\label{c12} p_{\varphi,eff}=\frac{1}{\kappa^2}\left[\kappa^2\bigg(\frac{1}{2}\dot\varphi^2-V\bigg)
+2H\dot{F}+\ddot{F}+ \chi \kappa^2 \varphi^2(3H^2+2\dot{H})-\xi \frac{3\dot F^2}{4F}\right].\;\ee
Thus, the effective equation of state   of  dark energy can be  expressed as
\be\label{c14} w_{DE}(z)=\frac{\kappa^2\bigg(\frac{1}{2}\dot\varphi^2-V\bigg)
+2H\dot{F}+\ddot{F}+ \chi  \kappa^2 \varphi^2(3H^2+2\dot{H})-\xi \frac{3\dot F^2}{4F}}{\kappa^2\bigg(\frac{1}{2}{\dot\varphi}^2+V\bigg)
-3H\dot F- 3\chi  \kappa^2H^2\varphi^2 -\xi \frac{3\dot F^2}{4F}}.\;\ee
Varying the action [Eq.~(\ref{action1})] with respect to the scalar field $\varphi$, we obtain  the modified Klein-Gordon equation~\cite{Baccigalupi, Amendola}
\be\label{c13} \ddot\varphi+3H\dot\varphi+V_{,\varphi}-\frac{F_{,\varphi}}{2\kappa^2} \hat R=0,\;\ee
where $V_{,\varphi}=\frac{d V}{d \varphi}$ and $F_{,\varphi}=\frac{d F}{d \varphi}$.

\begin{figure}[!htbp]
\centering
\includegraphics[width=9cm,height=6cm]{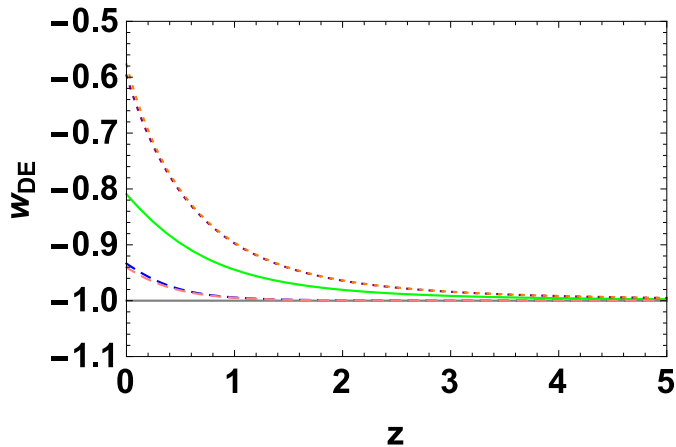}
\caption{\label{Fig1}Evolutionary curves of the equation of  state of dark energy  $w_{DE}$   with respect to the redshift $z$ for different  models. The gray and green solid lines correspond to the models of the cosmological constant  and minimally coupled quintessence, respectively. The blue and pink dashed lines represent the nonminimally coupled quintessence in the metric and Palatini formalisms, respectively, with $\chi=-0.5$,  while the purple and orange dotted lines correspond to them, respectively, with $\chi=0.15$.}
\end{figure}

Now we study the differences among different dark energy models in terms of the equation of state,  including the cosmological constant, and both minimally and nonminimally coupled quintessences, through  numerical calculations. Figure~\ref{Fig1} shows the evolution of the equation of  state of dark energy $w_{DE}$ with respect to the redshift $z$. When  $z\geq5$, all dark energy models have the same state parameter which approaches $-1$.
As the role of the dark energy becomes important  in the low-redshift region, the effect of the nonminimal coupling becomes apparent. In both formalisms,
a positive nonminimal coupling leads to values of $w_{DE}$  larger than those in the minimal coupling case in the redshift region $z<5$, while a negative nonminimal coupling yields smaller values. However, the difference between the two different formalisms is very small.
For the positive coupling case, the value of $w_{DE}$ in the Palatini formalism is slightly larger than that in the metric formalism, and an opposite result appears for the negative case.

\begin{figure}[!htbp]
\centering
\includegraphics[width=7.cm,height=5.45cm]{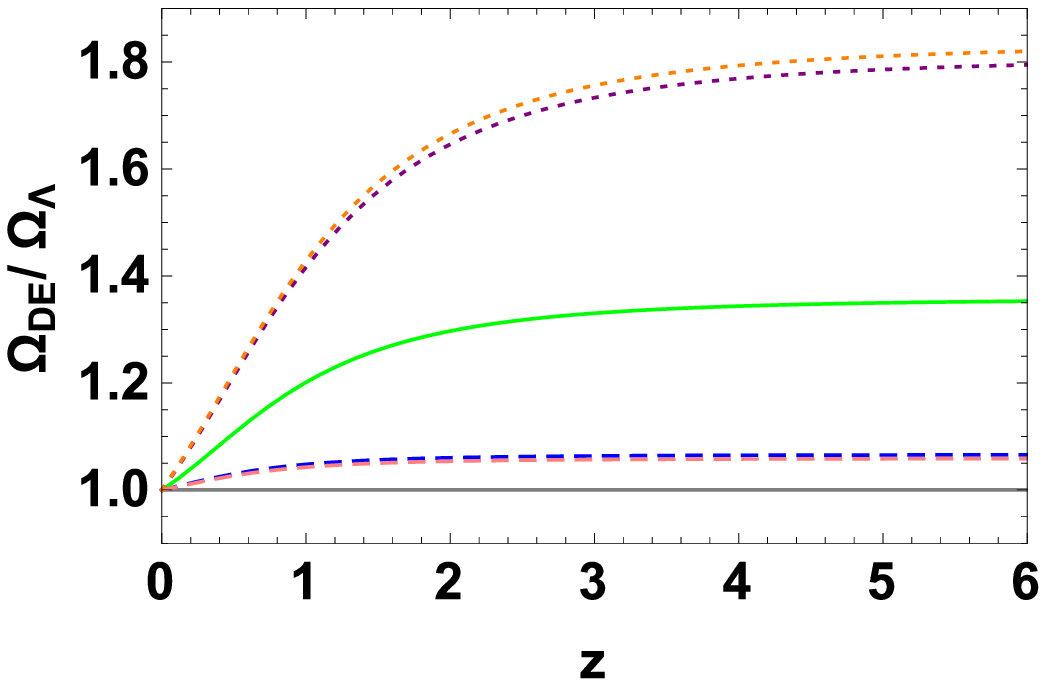}
\includegraphics[width=7.cm,height=5.45cm]{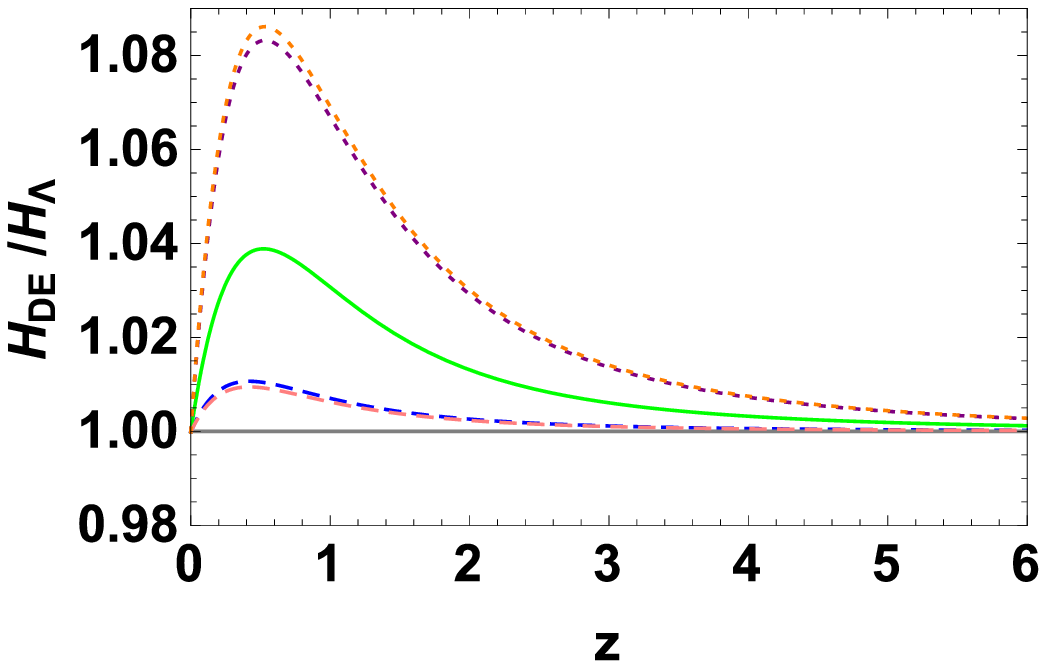}
\caption{\label{Fig2}Evolutions of the ratio of the energy density of dark energy (left panel) and the Hubble parameter (right panel) for the
extended quintessence models to their counterparts in the $\Lambda$CDM model. Line types  are the same as those described in Fig.~\ref{Fig1}.}
\end{figure}

Figure~\ref{Fig2} shows the evolutions of the ratio of the energy density of dark energy (left panel) and the Hubble parameter (right panel) to their counterparts in the $\Lambda$CDM model.
From the left panel, one can see that the Palatini formalism  with a   positive  coupling constant  leads to the greatest deviation from the $\Lambda$CDM, while the metric formalism with a negative   coupling constant yields  the least, and this indicates that  the value of the coupling constant has a significant effect on the energy density of dark energy.
The difference between  the metric and Palatini formalisms becomes apparent when $z>2$.
Turning to the right panel, we find that the differences among different dark energy models are small for the Hubble parameter, and the greatest  deviation from the $\Lambda$CDM is about $8.5\%$.
The small differences in the Hubble parameter indicate that the differences in the comoving and luminosity distance are small, and the different cosmological models give almost the same background evolution history.

\section{Perturbation equations}
We have studied the linear perturbations using the quasi-static approximation  in Ref.~\cite{Fan2015}. Now, we take a further step to  the nonlinear perturbations. With the assumption that the dark energy does not clump, one can start with
the continuity and Euler equations of the pressureless perfect fluid, which read~\cite{Amendola2010,Tatekawa2005,Peebles1981}
\be\label{c15}
\left(\frac{\partial \rho_{m}}{\partial t}\right)_{r}+\nabla_{r} \cdot (\rho_{m} \emph{\textbf{u}})=0,\;\ee
\be\label{c16} \left(\frac{\partial \emph{\textbf{u}}}{\partial t}\right)_{r}
+(\emph{\textbf{u}} \cdot \nabla_{r})\emph{\textbf{u}}=-\nabla_{r} \Psi_{N},\;\ee
where $\emph{\textbf{u}}$ is its moving velocity,
and $\Psi_{N}$ is the Newton gravitational potential.
The subscript $r$ indicates that   the physical coordinate $\emph{\textbf{r}}$ is used.  For convenience, we now transform the physical coordinate to the comoving one,
$\emph{\textbf{x}}=\emph{\textbf{r}}/a(t)$. Thus, we have  \be\label{c017} \emph{\textbf{u}}=\dot a \emph{\textbf{x}}+\emph{\textbf{v}}(\emph{\textbf{x}},t),\;\ee
where $\emph{\textbf{v}}=a \dot {\emph{\textbf{x}}}$ is the comoving peculiar velocity.
The energy density $\rho_{m}$ of pressureless matter, when the perturbation is considered,  can be expressed as
\be\label{c019} \rho_{m}(\emph{\textbf{x}},t)=\bar \rho_{m}(t)(1+\delta (\emph{\textbf{x}},t)),\;\ee with
the background energy density $\bar \rho_{m}(t)$ satisfying  $\bar \rho_{m}(t)\propto a^{-3}$ and
$\delta(\emph{\textbf{x}},t)\equiv \delta \rho_{m}(\emph{\textbf{x}},t)/\bar \rho_{m}(t)$ being the density contrast.

Using the transformation~\cite{Amendola2010,Tatekawa2005,Peebles1981}
\be\label{c20} \nabla_{x}=a \nabla_{r}, \quad \left(\frac{\partial y(\emph{\textbf{x}},t)}{\partial t}\right)_{r}=\left(\frac{\partial y}{\partial t}\right)_{x}-\frac{\dot a}{a}(\emph{\textbf{x}} \cdot \nabla_{x})y,\;\ee
with $y$ being an arbitrary function of $\emph{\textbf{x}}$ and $t$, and
substituting Eqs.~(\ref{c017}) and (\ref{c019})  into Eq.~(\ref{c15}), we obtain   \be\label{cc2} \frac{\partial\delta}{\partial t}+\frac{1}{a}\nabla_{x}\cdot(1+\delta)\emph{\textbf{v}}=0.\;\ee
 Here  the relations $H \nabla_{x}\cdot(\delta \rho_{m} \emph{\textbf{x}})=
H (\emph{\textbf{x}}\cdot\nabla_{x})\delta \rho_{m}+3H \delta \rho_{m}$
and $\dot{\bar\rho}_{m}+3H\bar \rho_{m}=0$ are used.

Combining Eqs.~(\ref{c017}) and (\ref{c20}) gives~\cite{Amendola2010}
\be\label{cc3} \left(\frac{\partial \emph{\textbf{u}}}{\partial t}\right)_{r}=\ddot a \emph{\textbf{x}}
+\left(\frac{\partial \emph{\textbf{v}}}{\partial t}\right)_{x}-H\dot a \emph{\textbf{x}}-H(\emph{\textbf{x}}\cdot \nabla_{x})\emph{\textbf{v}},\;\ee
\be\label{cc4} (\emph{\textbf{u}}\cdot \nabla_{r})\emph{\textbf{u}}=H\dot a \emph{\textbf{x}}+H \emph{\textbf{v}}
+H(\emph{\textbf{x}}\cdot \nabla_{x})\emph{\textbf{v}}+\frac{1}{a}(\emph{\textbf{v}}\cdot \nabla_{x})\emph{\textbf{v}}.\;\ee
Inserting Eqs.~(\ref{cc3}) and (\ref{cc4}) into Eq.~(\ref{c16}), we obtain the perturbation equation in the comoving coordinate
\be\label{cc5} \bigg( \frac{\partial \emph{\textbf{v}}}{\partial t}\bigg)_{x}+H \emph{\textbf{v}}
+\frac{1}{a} \emph{\textbf{v}}\cdot \nabla_{x} \emph{\textbf{v}}=\frac{1}{a}\nabla_{x} \Psi,\;\ee
where  $ \Psi\equiv-\Psi_{N}-\frac{1}{2}a \ddot a \emph{\textbf{x}}^2$, which satisfies  the Poisson equation \be\label{cc7} \nabla^{2} \Psi=4\pi G_{eff}\bar \rho_{m} a^{2} \delta,\;\ee
with $G_{eff}$ being the effective Newton constant and $G_{eff}=G$ in the general relativity limit.
For the extended quintessence models, the effective Newton constant has the form~\cite{Fan2015}
\be\label{c026} G_{eff}=\frac{G}{F}\frac{2F-3\xi F_{,\varphi}^2+4F_{,\varphi}^2}{2F-3\xi F_{,\varphi}^2+3F_{,\varphi}^2}.\;\ee

In the following, for simplicity,   the subscript $x$ is dropped. In addition, we rewrite Eqs.~(\ref{cc2}) and (\ref{cc5}) with the velocity vector $v_{i}$,  and obtain
\be\label{cc8} a {\partial \delta\over \partial t}=-\nabla^{i} \left[(1+\delta)v_{i}\right],\;\ee
\be\label{cc9} a {\partial   v_{i} \over \partial t} =-a H v_{i}-v_{j}\nabla^{j}v_{i}+\nabla_{i} \Psi.\;\ee
As the perturbation variables depend on space and time, their derivatives satisfy
\be\label{c25} \dot \delta(\emph{\textbf{x}},t)= {\partial \delta\over \partial t} +\frac{1}{a}v^{i} \nabla_{i} \delta
=-\theta(1+\delta),\;\ee
\be\label{c26} \dot \theta(\emph{\textbf{x}},t) =-2H \theta-\frac{1}{3}\theta^2-4\pi G_{eff} \bar \rho_{m}\delta.\;\ee
Here, we use the relation
$ \nabla^{i}(v_{j} \nabla^{j} v_{i})=(\nabla^{i} v_{j})(\nabla^{j} v_{i})+v_{j}\nabla^{j}\nabla^{i}v_{i}$, and the assumption that the peculiar velocity in the initial collapse region only relates to the radius so that~\cite{Amendola2010}
\be\label{c028} (\nabla^{i} v_{j})(\nabla^{j} v_{i})=\frac{1}{3}a^{2}\theta^2.\;\ee

Taking the derivative of Eq.~(\ref{c25}) with respect to time $t$ and inserting Eq.~(\ref{c26}) into it, we obtain the nonlinear perturbation equation for the pressureless perfect
fluid
\be\label{c27} \ddot \delta+2H \dot \delta-\frac{4}{3}\frac{\dot \delta^2}{1+\delta}-4\pi G_{eff} \bar \rho_{m} \delta(1+\delta)=0,\;\ee
and the corresponding   linear equation
\be\label{c28} \ddot \delta+2H \dot \delta-4\pi G_{eff} \bar \rho_{m}\delta=0.\;\ee

\section{Spherical collapse}
In this section, we study the spherical collapse  of pressureless matter  on the basis of the knowledge about the nonlinear perturbations. Supposing that there is a shell of dark matter at distance $\mathcal{R}$ from the center of a spherical overdensity,
and that the energy density  $\rho_{m}(\emph{\textbf{x}},t)$  of dark matter is always homogeneous in the spherical region,  one can find that  $\rho_{m}(\emph{\textbf{x}},t)$ satisfies the relation
\be\label{c30} \dot \rho_{m}+3h \rho_{m}=0,\;\ee
where $h=\dot{\mathcal{R}}/\mathcal{R}$ represents the local expansion rate inside the spherical region. At the same time, for the background dark matter, its energy density $\bar \rho_{m}(t)$ remains to satisfy the  continuity equation
\be\label{c29} \dot{\bar\rho}_{m}+3H\bar \rho_{m}=0.\;\ee
According to the assumption that the dark energy does not cluster, we have that the second Friedmann equations in the background and the spherical regions are
\be\label{c31} \frac{\ddot{a}}{a}=-\frac{4\pi G}{3}[\bar \rho_{m}+(1+3 w_{DE})\rho_{\phi,eff}],\; \quad
\frac{\ddot{\mathcal{R}}}{\mathcal{R}}=-\frac{4\pi G}{3}[ \rho_{m}+(1+3 w_{DE})\rho_{\phi,eff}].\;\ee
If one defines the density contrast as
\be\label{c031} \delta(\emph{\textbf{x}},t)=\frac{\rho_{m}}{\bar \rho_{m}}-1,\;\ee
inside the shell and $ \delta(\emph{\textbf{x}},t)=0$ outside,  $\delta(\emph{\textbf{x}},t)$ is a top-hat function, and it satisfies the same nonlinear and linear equations as those  [Eqs.~(\ref{c27}) and (\ref{c28})] derived for a shear-free fluid. Therefore, an initially spherical perturbation remains spherical.

Solving numerically Eqs.~(\ref{c27})$-$(\ref{c31}), one can see that the spherical region expands and the perturbation density grows with the expansion of the Universe at early times. When the spherical region expands to its maximum radius $\mathcal{R}_{ta}$ at $z_{ta}$, the peculiar velocity becomes
zero at this moment; then the expansion turns around, the overdensity turns to collapse, and the radius reduces and finally  reaches a singularity ($\mathcal{R}=\mathcal{R}_{c}\rightarrow0$) at $z_{c}$. However, since the dust assumption will fail at some high density and nonradial fluctuations will develop, the final singular phase is of course unphysical. The system
will finally settle into the virial equilibrium and stabilizes at the virialization radius $\mathcal{R}_{vir}$.

Since the linear density contrast  $\delta_{c}$  at collapse time $z_{c}$ is  used in the Press-Schechter theory as a first approximation to the epoch of galaxy formation and to the calculation of the abundance of collapsed objects, it connects with the observations and thus is an interesting parameter to be investigated first.  We  solve the nonlinear equation~(\ref{c27}) to determine $z_{c}$, which depends on the amplitude of the initial perturbation. The larger $z_{c}$ is, the earlier the overdense region will collapse. $\delta_{c}$ can be achieved by  solving  Eq.~(\ref{c28}) with the same initial conditions. Varying the initial condition to give different collapse redshifts $z_{c}$, we then obtain a redshift-dependent expression for the critical density, $\delta_{c}=\delta_{c}(z_{c})$.  In our calculation we adopt the cosmological parameter $\Omega_{m0}=0.3$  except for  the EdS model,   keep the initial conditions $\delta_{i}=3\ast10^{-4}$ and $\dot \delta_{i}=0$ fixed, and vary the initial redshift $z_{i}$.

\begin{figure}[!htbp]
\centering
\includegraphics[width=9cm,height=6cm]{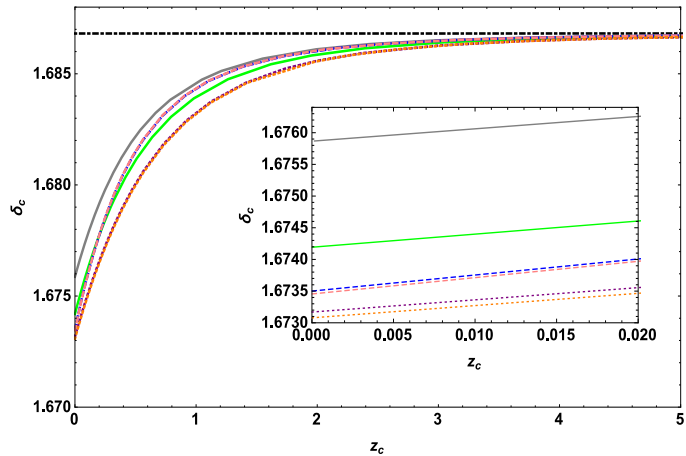}
\caption{\label{Fig4}Evolutions of the linear density contrast $\delta_{c}$ with respect to the redshift $z_{c}$ for different cosmological models. The black dot-dashed, gray solid, and green solid lines represent the Einstein$-$de Sitter, $\Lambda$CDM, and minimally coupled quintessence models, respectively. The blue and pink dashed lines represent the nonminimally coupled quintessences in the metric and Palatini formalisms, respectively, with $\chi=-0.5$,  while the purple and orange dotted lines correspond to them, respectively, with $\chi=0.15$.}
\end{figure}

Figure~\ref{Fig4} shows the evolutions of $\delta_{c}$ for different cosmological models.  For the EdS model,  $\delta_{c}$  is a constant and equals $1.6868$, which is shown as  the black dot-dashed line in Fig.~\ref{Fig4}.
One can see that all models converge to the EdS  at the high redshift, since the dark energy can be neglected and  the effective gravity constant $G_{eff}\approx G$ deep in the matter-dominated era~\cite{Fan2015}. For minimally and nonminimally coupled quintessence models, the values of $\delta_{c}$ are always less than that of $\Lambda$CDM.  In the very low-redshift region,  the values of $\delta_{c}$ for nonminimally coupled quintessence are less than that of the minimal one, which is independent of the value of the coupling constant and the formalism; although, when $z$ is larger than  about $0.3$, $\delta_{c}$ for nonminimally coupled models with a negative coupling is larger than that for the normal quintessence model. For the extended quintessence  with a positive coupling, the values of $\delta_{c}$ are always less than those  with a negative coupling.
 Furthermore,  different metric formalisms have negligible effects on $\delta_{c}$. But
the metric formalism gives a slightly larger    $\delta_{c}$ than  the Palatini formalism, and in the positive nonminimal coupling case the  differences between two different formalisms  are more obvious than that in the negative nonminimal couping.

As mentioned above, the final phase of the spherical collapse will be the virial equilibrium state, in which  the kinetic energy $T$ and the gravitational potential energy $U$ satisfy the relation
\be\label{c33} T=\frac{R}{2}\frac{\partial U}{\partial R}.\;\ee
For an inverse-power potential, the virialization implies $T=-U/2$. As the dark energy does not cluster, it just modifies the background
evolution of the system.  Thus, the energy density of dark energy in the cluster region is time-varying  with the same behavior as its background density,  which leads to 
the energy within the cluster region not being conserved. However, since the energy density of  dark matter  in the cluster region is much larger than the background one,  we can neglect the variation of the background density  and assume that the energy conservation law is still effective to study the virialization.

It is well known that the self-energy $U_{mm}$ of a sphere of nonrelativistic particles with mass $M$ ($M=4\pi \rho_{m} \mathcal{R}^3/3$) and radius $\mathcal{R}$ is \be\label{c34} U_{mm}=-\frac{3}{5}\frac{G M^2}{\mathcal{R}}.\;\ee
In the study of the evolution of spherical overdensities in the presence of homogeneous extended quintessence dark energy,  the gravitational effects of dark energy on dark matter can be taken into account by replacing $G$ with $G_{eff}$ in Eq.~(\ref{c34}),  yielding an addition term $U_{mD}$ on the gravitational potential energy of the spherical dark matter overdensity.  Since the scalar field only modifies the background,  we assume that the traditional recipes in the literature to evaluate the virial overdensity are still valid \cite{Wang98}, and then
\be\label{c35}   U_{mD}=-\frac{4\pi G_{eff}}{5}M \rho_{DE} \mathcal{R}^2.\;\ee
The conservation of energy at turnaround and at the virialization give
\be\label{c35b} T_{ta}+U_{ta}=T_{vir}+U_{vir}.\;\ee
Since $T_{ta}=0$,  inserting Eqs.~(\ref{c33}), (\ref{c34}), and (\ref{c35}) into Eq.~(\ref{c35b}), we get
\be\label{c36} U_{mm,ta}+U_{mD,ta}=\frac{1}{2}U_{mm,vir}+2U_{mD,vir}.\;\ee
Defining the spherical collapse parameter we are going to study as the overdensity at virialization
$\Delta_{v}=\rho_{m,vir}/\bar \rho_{m,vir}$, and assuming the system virialized at $z=z_{c}$,   we can obtain the following relation~\cite{Wang98}:
\be\label{c37} \Delta_{v}(z=z_{c})=\zeta {\left(\frac{\mathcal{R}_{ta}}{\mathcal{R}_{vir}}\right)}^{3} {\left(\frac{1+z_{ta}}{1+z_{c}}\right)}^{3}.\;\ee
Here $\zeta=\rho_{m,ta}/\bar \rho_{m,ta}$ is  the overdensity at turnaround.
Inserting Eqs.~(\ref{c34}) and (\ref{c35}) into Eq.~(\ref{c36}) and using the definitions of $\delta_{c}$ and $\Delta_{v}$, we arrive  at the equation of the variable
${\mathcal{R}_{vir}}/{\mathcal{R}_{ta}}$:
\be\label{c38} \left[1+\frac{1-\Omega_{m}(z_{ta})}{\zeta \Omega_{m}(z_{ta})}\right]\frac{\mathcal{R}_{vir}}{\mathcal{R}_{ta}}
-\frac{2}{\zeta}\frac{1-\Omega_{m}(z_{vir})}{\Omega_{m}(z_{vir})}
{\left(\frac{1+z_{c}}{1+z_{ta}}\right)}^{3}{\left(\frac{\mathcal{R}_{vir}}{ \mathcal{R}_{ta}}\right)}^{3}=\frac{1}{2}.\;\ee
For the EdS model, it is easy to see that   ${\mathcal{R}_{vir}}/{\mathcal{R}_{ta}}=1/2$, since $\Omega_{m}=1$. Supposing that the solution of Eq.~(\ref{c38}) has the form
${\mathcal{R}_{vir}}/{\mathcal{R}_{ta}}=1/2+\varepsilon$($\varepsilon$ is a small variable)~\cite{Lahav1991}, inserting it back into Eq.~(\ref{c38}) and dropping out the higher-order terms, we obtain
the approximate solution~\cite{Wang98}
\be\label{c39} \frac{\mathcal{R}_{vir}}{\mathcal{R}_{ta}}=\frac{1-\eta_{vir}/2}{2+\eta_{ta}-3\eta_{vir}/2},\;\ee
where
\be\label{c40} \eta_{ta}\equiv \frac{2}{\zeta}\frac{1-\Omega_{m}(z_{ta})}{\Omega_{m}(z_{ta})}, \quad
\eta_{vir}\equiv \frac{2}{\zeta}\frac{1-\Omega_{m}(z_{vir})}{\Omega_{m}(z_{vir})}
{\left(\frac{1+z_{c}}{1+z_{ta}}\right)}^{3}.\;\ee

\begin{figure}[!htbp]
\centering
\includegraphics[width=9cm,height=6cm]{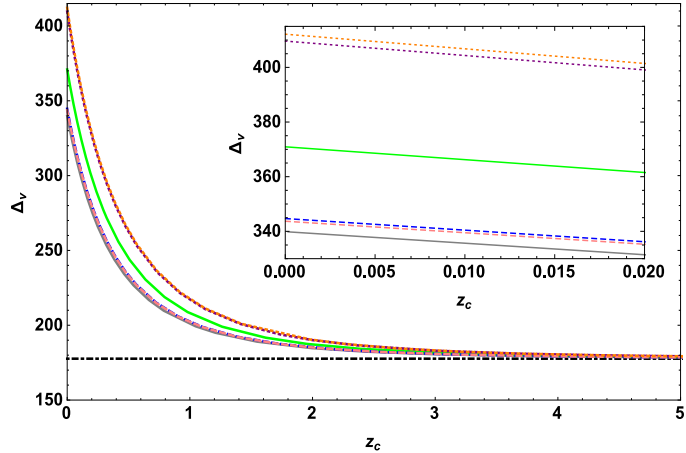}
\caption{\label{Fig5}Evolutions of the nonlinear overdensity $\Delta_{v}$ with respect to the redshift $z_{c}$ for different extended quintessence models. Line types and colors are as the same as those in Fig.~\ref{Fig4}.}
\end{figure}

Figure~\ref{Fig5} presents the evolutions of $\Delta_{v}$ for different cosmological models with respect to the redshift $z_{c}$.
All  models converge to the EdS at high redshift, which is the same as what happens for $\delta_{c}$. The values of $\Delta_{v}$ for all minimally and nonminimally coupled quintessence models are  larger than that of the $\Lambda$CDM model in the low-redshift region.
The nonminimally coupled models with a positive coupling constant  give larger values of $\Delta_{v}$ than  the  minimally coupling one, while the models with  a negative coupling lead to smaller values. As with the case of $\delta_{c}$, different formalisms have very small effects  on $\Delta_{v}$.
For the positive nonminimal coupling, the value of $\Delta_{v}$ in the metric formalism is slightly smaller than that in the Palatini formalism, which  is just the opposite for the negative nonminimal coupling.

\section{Conclusion}
In this paper, using the spherical collapse method, we have studied the nonlinear structure formation in cosmological models with  the extended quintessence as dark energy in the metric and Palatini formalisms. By numerical calculation,  we obtain  the evolutionary curves of the linear density contrast $\delta_{c}$ and the virial overdensity $\Delta_{v}$. For both formalisms, when the coupling constant is negative, the deviations of $\delta_{c}$ and  $\Delta_{v}$ from their counterparts in $\Lambda$CDM  are less than those in the cases of minimal coupling and positive nonminimal coupling, and it is just the opposite for the models with a positive coupling constant. For the extended quintessence cosmological model with a negative coupling,  since the deviations are  less than $1\%$, all quantities dependent on $\delta_{c}$ or $\Delta_{v}$ are essentially
 unaffected if the linear density contrast or the virial overdensity of the $\Lambda$CDM model is used as an approximation. Furthermore, we found  that  the differences between different metric formalisms are very small and can be neglected, which  indicates that  the nonlinear structure formations cannot be used to distinguish the metric and Palatini formalisms.

\begin{acknowledgments}
We thank Prof. Lixin Xu for his useful discussions. This work was supported by the National Natural Science Foundation of China under Grants No. 11175093, No. 11222545, No. 11435006, and No. 11375092;  and by the  Specialized Research Fund for the Doctoral Program of Higher Education under Grant No. 20124306110001.
\end{acknowledgments}

\end{document}